\documentclass[twocolumn,aps,prc,floatfix,showpacs]{revtex4-1}
\usepackage{amsfonts}
\usepackage{amssymb}
\usepackage{amsmath}
\usepackage{amssymb}
\usepackage{graphicx}
\usepackage{dcolumn}
\usepackage{float}
\usepackage{hyperref}
\usepackage{appendix}
\usepackage{enumitem}
\begin{document}

\preprint{HEP/123-qed}
\title{ Better insight into the Strutinsky method (published version) }
\author{B. Mohammed-Azizi}
\email{aziziyoucef@gmail.com}
\affiliation{University of Bechar, Bechar, Algeria}
\date{\today }

\begin{abstract}
Strutinsky's method is reviewed through a new understanding. This method depends on
two free parameters: The smoothing parameter and the order of the curvature
correction. It turns out that this method is nothing but a compromise
between two fundamental conditions which are the so-called asymptotic limit
which comes from the so-called remainder which imposes a small as possible smoothing parameter and the smoothing
condition which forces that parameter to be, at least, slightly larger than the
inter shell spacing. In this paper, to find the best value of the smoothing
parameter, a new criterion is proposed instead of the plateau condition . This
new criterion is much more clear and free from ambiguities of the usual
plateau condition. It is also found, that the second free parameter, i.e.
the order of the curvature correction, plays an accessory role since, it is
connected intimately to the smoothing parameter, when the smoothing is realized. This paper provides a new and definitive insight into Strutinsky's method and its relationship with semi-classical methods.
\newline
Link to the published version: \url{https://link.aps.org/doi/10.1103/PhysRevC.100.034319}.
\end{abstract}

\maketitle

\section{Introduction}

In nuclear structure, the Hartree-Fock-Bogoliubov (HFB) method is the best choice to solve
the mean field approximation. In the 80's, because of the limited power of
computers, it was difficult to make such calculations. The use of
Strutinsky's method \cite{1966s,1967s,1968s} was then a good palliative.
This powerful method was particularly useful in the study of the binding energy and the fission barrier where
it obtained remarkable results \cite{1972br}. It was even difficult to
compete with it. Today, although less used than before, it continues to have
followers \cite{2018dobrov,2018ivan,2018qing,2018qingo}.

One of the weak points of this method is undoubtedly its inherent
ambiguities. First, Strutinsky's method is also called the
macroscopic-microscopic method because it associates two types of opposite
nuclear models and this is a priori not very coherent from the point of view of a fundamental theory. However, this
\textquotedblleft mixture\textquotedblright\ can be justified from the HFB
theory \cite{1981braquen}. Second, this method includes two free
parameters: The width of the smoothing functions and the order of the
curvature correction. It appears that the results of this method always
depend more or less strongly on these two parameters. In this respect, the plateau condition has been imposed to reduce this dependence. However, as we will see, this condition is not above criticism.
\newline
Actually, the most fundamental question is: what exactly is the Strutinsky
method? The precise response to this question has been given in 2006 in Ref. 
\cite{2006m} which shows that the Strutinsky method is only an
approximation of the semi-classical method \cite{1933k,1934w,1973j,1975j,1975jbb}. This approximation
contains an unavoidable remainder, which means that the method cannot give
an exact result, i.e. with the required precision. Once this has been
clarified, the problem of the method accuracy and its dependence on the two
free mentioned parameters remains to be solved. In this respect, this study
provides deeper analysis and further clarifications of the one which has
already been addressed in Reference \cite{2006m}. First, this paper
indicates, in particular, the way to obtain the optimal value for the
smoothing parameter. That is to say, for which the results are the closest
to those given by the semi-classical method. To this end, a basic criterion
is given (see Sec. \ref{sec:The-criterion-of}) in order to achieve this optimal value in a very precise manner. This criterion is a consequence of the established link.
Furthermore, this criterion which is fundamentally different from the
so-called plateau condition, will be justified in this paper. This study
will also explain why the order of the curvature correction has very little
influence on the results when the smoothing parameter is chosen optimally as long as the order is not too small.
When comparing the Strutinsky method with the semi-classical method, a number of remarks and questions come to mind. This study provides the appropriate answers to these problems.
This paper is the result of a very large number of numerical calculations
and checks. Among these calculations, I have chosen some examples to
illustrate the most important features. The numerical aspect has been
intensively processed using a large number of FORTRAN programs.
It is hoped that reading this paper will provide the necessary elements
that allows a good understanding of this method and the associated
ambiguities and will explain how to handle the Strutinsky procedure.

\section{Strutinsky method vs semi-classical method: Highlights} \label{green}

Before going into the details of the subject of this paper, it is necessary to draw a parallel between the Strutinsky method and the semi-classical method. This will help us to have an overview of both methods in order to compare them, to see what separates them from what brings them together. This section is dedicated to the salient features of these methods (see also Ref. \cite{2010bh} and references quoted therein). \newline In the microscopic-macroscopic method, the shell correction to the classical liquid drop model results from the quantum effects of the atomic nucleus.  This shell correction can be mainly calculated in two ways: (i) The first one uses the Strutinsky method, which was widely used in the past, (ii) The second uses one of the semi-classical methods which is generally the one based on the Wigner-Kirkwood $\hbar$  expansion \cite{1933k,1934w} up to the fourth order (Thomas-Fermi term plus  $\hbar^2$ and $\hbar^4$ corrections). The second method has been used fairly successfully recently \cite{2010bh}. \newline  It has been found in several studies that in most cases, the results of these two methods are very close \cite{1975jbb,1973b,1998ve,1976je}, though they are not exactly the same. This similarity between the two methods has been found in a large number of papers, but without establishing a direct link.
Both methods use level density as the basic quantity from which a number of particles and energy can be deduced. Furthermore, it has been early stated in Ref. \cite{2001ul} that the:"Strutinsky shell correction method is essentially a semi-classical approximation. It rests on the fact that the number of particles in the system considered is large, rather than on the interaction between the particles being weak". There are many other arguments in favor of a link between the two methods (see the rest of this text).
Thus, the fundamental question then is whether these densities are completely different, whether they are close or whether there is a link between them. \newline Before answering this fundamental question, it is useful to note that the two methods encounter different difficulties in practice. \newline In the Strutinsky method, the main difficulty lies in the treatment of the continuum for finite wells. In effect, this method works with the  discrete eigenvalues (bound states) and continuum. But, the latter is more difficult to obtain from the Schrodinger equation. The continuum plays an even more important role as the Fermi level is closer to it. This can be delicate for nuclei close to the drip line. Originally, in this case, the discrete levels of positive energies (the so-called quasi-bound states resulting from diagonalization in a finite harmonic oscillator basis) were artificially used to "simulate" the continuum. But other methods are possible. To my knowledge the most rigorous and easy treatment of the continuum is given by the Green functions method which in practice amounts to diagonalizing two Hamiltonians \cite{2000ve}. In other words, to apply the Strutinsky method it is necessary to solve Schrodinger's equation rigorously. \newline Another difficulty results from the plateau condition, which is not satisfied in all cases. This remark of the absence of a clear plateau in Strutinsky's method has been very often made in the literature. It turned out that even in the case where this treatment (continuum) is correctly performed, the plateau condition is seldom satisfied \cite{1998ve} (this remark is very important for this paper). An attempt to find an alternative to the plateau condition was already given in the same paper (see section \ref{sec:The-criterion-of}). \newline In the Wigner-Kirkwood method, the problem of the continuum does not exist since there is no explicit reference to this continuum. In addition, there is no restriction on whether the potential must be finite or not. Now, for finite wells (Wood-Saxon for example), one of the unfair criticisms \cite{1998ve} of this method is that the $\hbar^2$ correction in the semi-classical Wigner-Kirkwood density diverges in $\epsilon=0$ as $1/\sqrt{\epsilon}$  while for the Strutinsky density there is only a finite peak in $\epsilon=0$. It was then suggested that the Strutinsky method should be preferred. However, reference \cite{2007ce}  points out that the level density must be seen as a distribution in the mathematical sense of the term and that, for the number of particles as well as for energy, it appears only under the integral sign.  Consequently, although the level density diverges, these integrals are finite and therefore perfectly defined and this type of problem does not need to be addressed. Thus, in this respect, the Wigner-Kirkwood method was used to deduce a semi classical shell correction (with $\hbar^{2}$ and $\hbar^{4}$ corrections) without any problems of this type (see Ref. \cite{2010bh} and references quoted therein). However, from the point of view of the numerical aspect, the WK method remains a cumbersome procedure compared to the Strutinsky method. \newline The definitive answer to the fundamental question posed above, namely, the difference between the level density of the Strutinsky method and the semi classical method (Wigner-Kirkwood)  has been given by reference \cite{2006m}. The latter demonstrates that the Strutinsky density is only an approximation of the semi-classical method. Thus, in that paper \cite{2006m}, the so-called remainder explains why the results in the two methods are very close, but not rigorously the same. Unfortunately, some points of this demonstration do not seem to have been well understood. Therefore, I will take advantage of this paper to clarify as much as possible this demonstration so that there are no more ambiguities or doubts about it. First, we will clarify what is mean by the classical limit.

\section{ Bohr correspondence principle and classical limit} \label{cl}
The famous correspondence principle of large quantum numbers, first proposed by Niels Bohr in 1923, states that the quantum behavior of a system reduces to a classical physics, when the quantum numbers involved are very large.
\newline In a quantum system such as the atomic nucleus considered as a set of independent fermions  in an average field, the Fermi energy level determines the energy of the nucleus. The larger the number of nucleons, the higher the Fermi level. Thus the classical limit could be defined by the fact that the Fermi level should be very large (measured from the bottom of the wells $V_{0}$) compared to one quantum energy of the system (or compared to the zero point energy) referred to as $\hbar\omega $, i.e.:
$\lambda-V_{0}\gg \hbar\omega$. \newline
For this reason, semi-classical methods  are more applicable to heavy nuclei than to light nuclei. The ideal case for such systems is given by the (unphysical) limit $\lambda\rightarrow \infty$, in a such way that, level density (or energy) could be defined by an asymptotic expansion which is somewhat a behavior  in the vicinity of  infinity. Such series can be obtained, for example, by the Wigner-Kirkwood method or by the Euler Maclaurin formula or other methods. The first term, i.e., the main contribution of such expansion is the Thomas-Fermi approximation.  For example, in the case of the three dimensional harmonic oscillator, the semi classical level density is given by \cite{2006m}: $g_{sc}(\lambda)\approx(1/2\hbar\omega)[(\lambda/\hbar\omega)^{2}-1/4]$ \newline
Thus the limit $\lambda\rightarrow \infty$ becomes in practice $\lambda\gg\hbar\omega$, here, $V_{0}=0$.\newline
In the classical limit, all quantum quantities become close to their analogs in classical physics. In particular, the quantum level density [see in the following  Eq. (\ref{eq:one})] becomes close to the one deduced from the semi-classical method. Schematically, in the general case, we will have (with obvious notations):
\begin{equation}
g_{0}(\epsilon)\approx g_{sc}(\epsilon)
\label{equiv}
\end{equation}

in the classical limit:
\begin{equation}
 \lambda-V_{0}\gg\hbar\omega 
 \label{scl}
 \end{equation} 
 Within the limit of large quantum numbers, quantum mechanics becomes a classical mechanics. Consequently, in this limit, quantum effects disappear and shell effects that are quantum effects also disappear.

\section{Reminder on the principle of the microscopic-macroscopic method}
The principle of that method \cite{1972br} is based on the fact that the
energy of a nucleus can be split into a smooth part which varies slowly
with the number of neutrons and protons and a rapid fluctuations due to the
shell structure of the level density. The justification for such a
separation has been made on the basis of the Hartree-Fock theorem \cite%
{1981braquen}. \newline
The smooth part is generally deduced from a classical model such as the
liquid drop model and quantum (shell) corrections are derived from a microscopic
model. The present work concerns only the shell correction, the pairing
correction is therefore, outside of the scope of this study. Shell effects
are evaluated separately for neutrons and protons. In the Strutinsky method,
the binding energy is given by:
\begin{equation}
Energy\text{ = }E^{liquiddrop}\text{+}\delta E_{neut}^{shellc}\text{ +}%
\delta E_{prot}^{shellc}
\end{equation}%
Where each of the shell corrections is defined as:
\begin{equation}
\delta \overline{E}_{M,\gamma }=\sum_{n=0}^{NorZ}\epsilon _{n}-\overline{E}%
_{M,\gamma }(\lambda )  \label{eq:shellc-1}
\end{equation}%
in which the first sum (which contains shell effects) is that of the single-particle energy-levels and the
second is the smooth energy (which is free of shell effects)  defined through the Strutinsky procedure. $N$
and $Z$ are the neutron and proton numbers. The
Strutinsky procedure depends on two free parameters: The order $M$ and the
smoothing parameter $\gamma $.\newline
In this regard, it is worthwhile to note that the smooth energy generated by the Strutinsky method has recently been replaced by the one derived (semi-classically) from the Wigner-Kirkwood method (up to the order $\hbar^{4}$) which does not contain free parameters \cite{2010bh}, without any problem.

\section{The Strutinsky level density and the Strutinsky energy. Basic formulas}
\subsection{The exact or quantum level density}
The exact (or sharp or quantum) level density $g_{o}(\epsilon)$, which contains shell effects, for neutrons or protons
is defined as a sum of Dirac functions on the basis of the knowledge of the
set of energy levels $\left\{ \epsilon_{n}\right\} $.
\begin{equation}
g_{o}(\epsilon )=\sum_{n=0}^{\infty }\delta (\epsilon -\epsilon
_{n})=\sum_{n=0}^{\infty }\frac{1}{\gamma }\delta (\frac{\epsilon -\epsilon
_{n}}{\gamma })  \label{eq:one}
\end{equation}%

In Eq. (\ref{eq:one}) the parameter $\gamma $ is introduced to make the
argument of delta function, dimensionless. Subsequently, it will play the
role of the width of the Gaussian smoothing-functions (see below). 
\newline The energy interval between two successive shells
constitutes a shell gap. For a spherical nucleus, each gap is characterized by one of the well-known magic
numbers in the famous shell model of Maria/Haxel \cite{1949mayer,1949haxel}. But secondary gaps and other magic
numbers can appear for deformed nuclei, the so-called
deformed magic numbers \cite{1972br}.\newline
The delta functions (especially for high degeneracy of energy levels)
represent abrupt variations of the level density. Each delta function is
centered at an energy level $\epsilon _{n}$. \newline
Within the expression of Eq. (\ref{eq:one}) the energy (that contains the
shell effects) is defined by:
\begin{equation}
E=\int_{-\infty}^{\lambda_{0}}\epsilon
g_{o}(\epsilon)d\epsilon=\sum_{n=0}^{NorZ}\epsilon_{n}  \label{eq:two}
\end{equation}
Where $\lambda_{0}$ is the sharp Fermi level. In principle it is derived
from the condition of conservation of the number of particles:
\begin{equation}
NorZ=\int_{-\infty}^{\lambda_{0}}g_{o}(\epsilon)d\epsilon  \label{eq:three}
\end{equation}
Due to the fact that the exact level density is based only on delta (sharp)
functions [given by Eq. (\ref{eq:one})], the Fermi level $\lambda_{0}$ cannot be
completely determined. Generally, for a sharp distribution of nuclear
matter, it is defined as the highest occupied level. Nevertheless, here,
there is no need to know $\lambda_{0}$ because it does not appear explicitly
in the sum of Eq. (\ref{eq:two}).\newline
Note that the exact density defined using delta functions is not a function in the conventional sense. It rather must be considered as a distribution. For physical quantities, such as the number of particles or the energy,  this distribution appears only under the integral sign and these quantities remain finite.

\subsection{The smoothing or averaging functions in the Strutinsky method}
In order to smooth out the quantum density [Eq. (\ref{eq:one})], i.e. to remove the
shell effects, Strutinsky thought first to replace the delta functions by
pure Gaussian functions with finite width $\gamma $. Thus, the basic idea of
Strutinsky is to \textquotedblleft spread the delta functions over an
interval of finite length $\gamma $. To eliminate the shell oscillations,
the width $\gamma $ of these Gaussians must be at least equal to the mean
gap between two successive shells usually denoted by $\hbar \omega $ (in
reference to the typical example of the harmonic oscillator). It turned out
that such replacement was not so accurate and led Strutinsky to introduce
the so-called curvature correction. This consists of multiplying the
Gaussian  by a polynomial of order $M$. The origin of the curvature correction is mainly due to the case of the harmonic oscillator because it provided a remarkable improvement for this particular case and also helped to improve the plateau condition (see below) for the other cases. Thus, the smoothing
procedure amounts to replacing delta functions in Eq. (\ref{eq:one}) [and
therefore in Eq. (\ref{eq:two}) and (\ref{eq:three})] by smoothing functions as
follows:

\begin{equation}
\frac{1}{\gamma}\delta(\tfrac{\epsilon-\epsilon_{n}}{\gamma})\rightarrow
P_{M}\left(\tfrac{\epsilon-\epsilon_{n}}{\gamma}\right)\dfrac{1}{\gamma\sqrt{%
\pi}}\exp\left(-\left(\tfrac{\epsilon-\epsilon_{n}}{\gamma}\right)^{2}\right)
\label{eq:mimic}
\end{equation}
with the important smoothing condition that the parameter $\gamma$ must
be at least of the order of the inter shell spacing.
\begin{equation}
\gamma\succsim\hbar\omega\text{ \ \ \ \ \ \ \ \textit{(smoothing condition)}}
\label{eq:smootc}
\end{equation}
To condense the notation, the smoothing functions will be denoted as:
\begin{equation}
P_{M}\left(\tfrac{\epsilon-\epsilon_{n}}{\gamma}\right)\dfrac{1}{\gamma\sqrt{%
\pi}}\exp\left(-\left(\tfrac{\epsilon-\epsilon_{n}}{\gamma}%
\right)^{2}\right)=\dfrac{1}{\gamma}F_{M}\left(\tfrac{\epsilon-\epsilon_{n}}{%
\gamma}\right)  \label{eq:fm}
\end{equation}
Making the replacement given by Eq. (\ref{eq:mimic}), the exact density of Eq. %
\ref{eq:one} will become a convenient regular continuous function (as soon
as Eq. (\ref{eq:smootc}) is fulfilled) and will be defined as the Strutinsky
(i.e. smooth or average) level density, which is free of shell effects :
\begin{equation}
\overline{g}_{M\text{ , }\gamma}(\epsilon)=\sum_{n=0}^{\infty}\dfrac{1}{%
\gamma}F_{M}\left(\frac{\epsilon-\epsilon_{n}}{\gamma}\right)
\label{eq:strutg}
\end{equation}
In realistic calculations the mean inter-shell spacing is usually taken of
the order of
\begin{equation}
\hbar\omega\approx41A^{-1/3}
\label{realistic}
\end{equation}
where $A$ is the mass number. Note the similarity between Eq. (\ref{eq:one})
and Eq.(\ref{eq:strutg}). The polynomial $P_{M}$ in the smoothing functions
is defined by means of Hermite polynomials:
\begin{equation}
P_{M}\left(u_{n}\right)=\sum_{m=0}^{M}A_{m}H_{m}\left(u_{n}\right),\text{ \ }%
u_{n}=\frac{\epsilon-\epsilon_{n}}{\gamma}
\label{first}
\end{equation}
\begin{equation}
A_{m}=\tfrac{(-1)^{m/2}}{\left[ 2^{m}\left( m/2\right) !\right] }\text{\ if }%
m\text{ is even, }\,\,\,\,A_{m}=0\text{ \ if }m\text{ is odd}
\end{equation}%
 Here, the definition of the smoothing function differs slightly from that of the reference \cite{2006m}, the factor ($1/\gamma$) being here external.
 Using the Darboux-Christoffel formula (see appendix of Ref. \cite{2006m}), another form of this polynomial can be written as:
 \begin{equation}
 P_{M}(u_{n})=\frac{H_{M}(0)}{2^{M+1}M!}\frac{H_{M+1}(u_{n})}{u_{n}}
 \label{another}
 \end {equation}
 Here, $M$ is even and $H_{M}(0)$ is given by:
 \begin{equation}
 H_{M}(0)=(-1)^{M/2}\frac{M!}{(M/2)!}
 \end{equation}

By means of the Strutinsky density, in a similar way  to Eq. (\ref{eq:two}), the Strutinsky (or average or smooth) energy will be given by:
\begin{equation}
\overline{E}_{M,\gamma}(\lambda)=\int_{-\infty}^{\lambda}\epsilon\overline{g}%
_{M,\gamma}(\epsilon)d\epsilon
\end{equation}
The shell correction of the Strutinsky method is then given by: 
\begin{equation}
\delta\overline{E}_{M,\gamma}=E-\overline{E}_{M,\gamma}(\lambda)
\label{eq:shellc}
\end{equation}
Where $E$ is defined by Eq. (\ref{eq:two}). The Fermi level $\lambda$  is determined by
the conservation of the particle number:
\begin{equation*}
NorZ=\int_{-\infty}^{\lambda}\overline{g}_{M,\gamma}(\epsilon)d\epsilon.
\end{equation*}

The detailed formulas of $\overline{g}_{M,\gamma}$ and $\overline{E}_{M,\gamma}$
 are given in Ref.  \cite{1972b}. There is no need to give these formulas
which are only used for FORTRAN programming.

\section{Main defect of the Strutinsky method}
The main problem of Strutinsky's method is that, since the shell correction is a physical quantity, it should not depend on the two free mathematical parameters  which are  the Gaussian width, represented by
the smoothing parameter $\gamma $ and the order $M$ of the curvature
correction.  The same remarks holds for the Strutinsky density of states  $\overline{g}%
_{M,\gamma }$ or the Strutinsky energy $\overline{E}_{M,\gamma }$.\newline
This led to imposing the so-called plateau condition ensuring at least
"locally" the independence of the shell correction with respect to these two
parameters
\begin{eqnarray}
\left[ \frac{\partial \delta \overline{E}_{M,\gamma }(\lambda )}{\partial
\gamma }\right] _{\gamma \succsim \hbar \omega} &\approx &0
\label{plato1} \\
\left[ \frac{\partial \delta \overline{E}_{M,\gamma }(\lambda )}{\partial M}%
\right] _{\gamma \succsim \hbar \omega} &\approx &0  \label{plato2}
\end{eqnarray}%
Usually, the plateau condition is searched for a fixed order $M$ and the
second condition is in most cases not taken into account. 
 In practice, it is often difficult to locate accurately the plateau
because in many cases it does not appear very clearly. Consequently, in this
method, the uncertainties and ambiguities are always present
In fact, all the ambiguities of Strutinsky's method are related to this type of problem. I will discuss this issue in more detail later in this paper.
\begin{figure*}
\includegraphics[width=1.5%
\columnwidth,keepaspectratio]{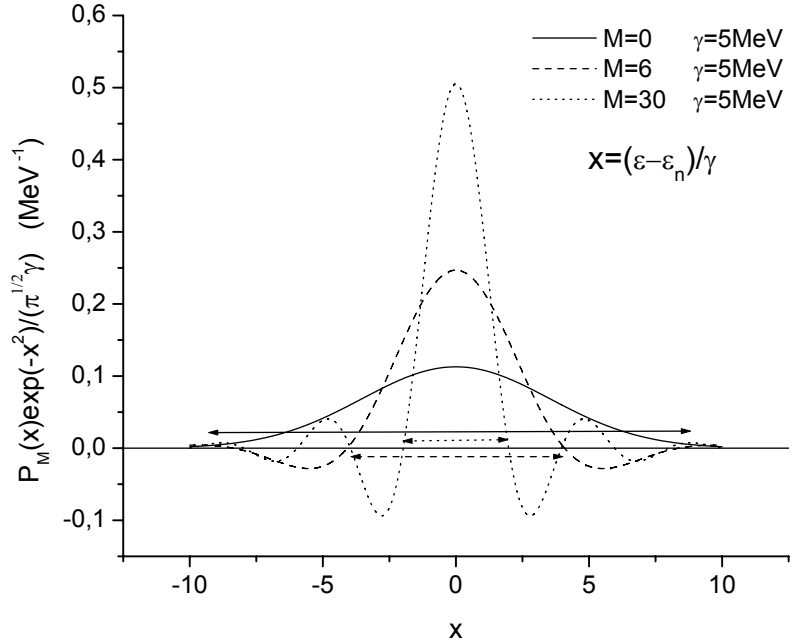}
\caption{ Smoothing functions Eq. (\protect\ref{eq:fm}) for three values of
the order $M$ and their width. The double arrow indicates roughly,  the width of the smoothing function, for each of the three cases ($M$=0; 6; 30).}
\label{figsmooth}
\end{figure*}

\section{Actual width of the smoothing functions depends not only on $\gamma$ but also on $M$ \label{sec:sed 5} }
Thus, the smoothing procedure of the Strutinsky method amounts to performing 
the replacement given by Eq. (\ref{eq:mimic}). In this procedure, each Dirac function in the sum of Eq. (\ref{eq:one}) is "mimed" by a continuous function (see Eq. (\ref{eq:fm})) with
a finite width. Unlike a sum of delta functions, the sum of modified
exponential gives as a result a continuous level density with oscillations.
The larger the width of these functions, the smaller the oscillations (fluctuations). \newline
The most important point concerns the real width of the smoothing
functions $F_{M}$. In the absence of the curvature correction the width of
the curve is the one of a pure Gaussian, i.e. represented by the sole
parameter $\gamma $. But the curvature correction, i.e. $P_{M}$ in Eq.(\ref%
{eq:fm}), must be taken into account. In effect, in that formula, we have a
product of a polynomial $P_{M}$ of order $M$ by a Gaussian. Therefore, it is
easy to see that the polynomial of the curvature correction influences the
width of the resulting curve. Consequently, the width of the smoothing
functions Eq. (\ref{eq:mimic}) will not be represented only by  the parameter  $\gamma$ as is often claimed. More precisely, the polynomial has $M$ roots
and vanishes $M$ times. Therefore, the first root defines practically the
true width of that curve. Thus, the real width of the smoothing
function depends not only on the parameter $\gamma$  but also on the order $M$.
 For this reason, when the order $M$ of the curvature correction increases, it is necessary to increase at the same time, the value of $\gamma$  so that the smoothing is actually 
achieved. Strictly speaking, the smoothing value must be indexed by $M$. \newline A more correct way of writing the smoothing condition would be: 
\begin{equation}
\gamma_{M}\succsim\hbar\omega 
\end{equation}
Which means that the smoothing value of the parameter $\gamma$ depends actually  on the order $M$. \newline
I illustrate in Fig. \ref{figsmooth} how the order $M$ modifies the smoothing functions, that is $(1/\gamma)F_{M}(x)=P_{M}(x)e^{-x^{2}}/(\pi^{1/2}\gamma)$ with $x=(\epsilon-\epsilon_{n})/\gamma$, when $\gamma$ is kept constant. 
One can compare the
width of the pure Gaussian ($M=0$) with two other curves, corresponding to 
$M=6$ and $M=30$. From that figure, it is clear that the real width of the
smoothing function depends also on the order $M$ and diminishes as $M$
increases. Thus for a fixed $\gamma $ the width decreases as $M$ increases.
It amounts to the same thing to increase $M$ or to diminish $\gamma $. This
explains why the smoothing functions reduce to a delta function for $\gamma
=0$ or $M=\infty $ and therefore why the Strutinsky level density given by
Eq. (\ref{eq:strutg}) reduces to the exact level density in these two distinct limit
cases. Now, it seems obvious, that $M$ and $\gamma$ are closely dependent in
the smoothing procedure.  This is why the plateau (or extremum) is moved to the right as $M$ increases. \newline
In this respect, it has already been noticed in reference \cite{1998ve} that, where the plateau condition was approximately satisfied,  a strong correlation is found between the values of $\gamma$ and $M$. This will be even clearer in section \ref{interrogation} in tables \ref{inf}, \ref{jnf} and \ref{knf} where we can clearly see that the $\gamma$ value that smoothes Strutinsky's density increases with the order $M$.

\section{Connection between the semi-classical level density and the
Strutinsky level density. Clarifications \label{sec: connection}}

In 2006, Ref. \cite{2006m} proved analytically that the shell correction
evaluated from the Strutinsky method is only an approximation of the one
deduced from the semi-classical method. The demonstration of the fundamental formula (24) of Ref. \cite{2006m} contains too many details, and could be difficult to understand at first sight. So, it seems useful, to clarify its main steps.

\subsection{The principle of the demonstration of this formula}
The Strutinsky level density can be obtained by means of the well known usual folding procedure of the quantum level density \cite{2006m}:
\begin{equation}
\overline{g}_{_{M},\gamma}(\lambda)=\int\limits_{-\infty}^{\infty}g_{o}(%
\epsilon)\frac{1}{\gamma}F_{M}\left( \frac{\epsilon-\lambda}{\gamma}\right) d\epsilon
\label{folding}
\end{equation}

With the smoothing condition (\ref{eq:smootc}). Here, the quantum level $g_{0}$ density is given by Eq. (\ref{eq:one}), $F_{M}$ is the smoothing function defined by Eq. (\ref{eq:fm}) and $\lambda$ is the Fermi level.
\newline It should be noted that the expression (\ref{eq:one}) is, in principle, valid only for infinite wells, even if for finite wells, the positive  energies (quasi-bound states) are sometimes used to simulate the continuum for finite wells (see section (\ref{green})). Strictly speaking, for a finite well, the expression (\ref{eq:one}) should be:
\begin{equation}
g_{0}(\epsilon)=\sum_{n}\delta (\epsilon -\epsilon_{n})+g_{c}(\epsilon)
\end{equation}
where $g_{c}(\epsilon)$ stands for the expression of the continuum. \newline
It is assumed that the resolution of the Schrodinger equation has been made and that the entire spectrum (discrete and continuous parts) is known. In this case, $g_{0}(\epsilon)$ is also completely known. The way in which the spectrum of eigenvalues has been resolved is of little importance in the demonstration since the quantum density only intervenes in a purely formal way in Eq. (\ref{folding}). Consequently, the demonstration of the fundamental relationship between the two types of level density is valid for both infinite and finite wells. \newline
Now, the most important point is based on the condition of the classical limit. Therefore, if we impose the \textbf{classical limit} given by Eq. (\ref{scl}) in Eq. (\ref{folding}), then the exact density (or quantum density) becomes close to the classical density, i.e. becomes the semi-classical density (see Eq. (\ref{equiv})), which is free of shell effects. We obtain thus, the basis of the demonstration:

\begin{equation}
\overline{g}_{_{M},\gamma}(\lambda)\approx\int\limits_{-\infty}^{\infty}g_{sc}(%
\epsilon)\frac{1}{\gamma}F_{M}\left( \frac{\epsilon-\lambda}{\gamma}\right) d\epsilon
\label{500}
\end{equation}
where $g_{sc}(\epsilon)$ is the semi-classical level density, with the condition of the semi-classical limit: $\lambda-V_{0}\gg\hbar\omega$, and the smoothing condition: $\gamma\gtrsim\hbar\omega$. \newline
Now, making $X=\dfrac{\epsilon-\lambda}{\gamma}$, we obtain:
\begin{equation}
\overline{g}_{_{M},\gamma}(\lambda)\approx\int%
\limits_{-\infty}^{\infty}g_{sc}(\lambda+\gamma X){F_{M}}\left(X\right) ) dX
\label{600}
\end{equation}
$M$ being even.

The next step is to perform a Taylor expansion of the semi-classical density around $\lambda$, up to the order ($M+2$), neglecting all the other terms. This means that the semi-classical density is approximated by the Taylor polynomial of degree ($M+2$).
\begin{equation}
g_{sc}(\lambda+\gamma X)\approx g_{sc}(\lambda)+\sum\limits_{m=1}^{M+2} \frac{(\gamma X)^m}{m!}\frac{d^mg_{sc}(\lambda)}{d\lambda^m}
\label{666}
\end{equation}
The first term, i.e. $g_{sc}(\lambda)$ is a constant in the integral so that the orthogonality property of Hermite polynomials shows that only the constant term $m=0$ in the first form of $P_{M}(x)$  (in Eq. (\ref{first}))  has a contribution (equal to the unity) so that the integral gives back to  i.e. $g_{sc}(\lambda)$. 
In the sum, the terms  $m=1,2,....(M+1)$ have zero contribution to the integral  (\ref{600}). This is easily seen, if we use the form given by Eq. (\ref{another}) for $P_{M}(x)$ employing the fact that the Hermite polynomial $H_{M}(x)$ is orthogonal to any polynomial of a lower degree.
  It is the last term ($M+2$) of this polynomial which leads to the so-called remainder. The latter is the cause of the uncertainty of Strutinsky's method. The last term gives an order of magnitude of the rest which is called here $R_{M+2,\gamma}$ instead of $R_{M,\gamma}$ as in Ref.  \cite{2006m}. In the mathematical demonstration, the remainder is obtained by integrating Eq. (\ref{600}) (see details in the original Ref. \cite{2006m}).  The result is:
\begin{equation}
\overline{g}_{_{M},\gamma}(\lambda)\approx g_{sc}(\lambda)(1+\,Remainder)
\label{rel}
\end{equation}
which shows that connection between the two level densities.

The order of magnitude of this remainder being given by \cite{2006m} (an exact expression will be given just below):
\begin{equation}
Remainder=R_{M+2,\gamma}(\lambda)\approx\frac{C_{M+2}\gamma^{M+2}}{g_{sc}(\lambda)}\left(%
\dfrac{d^{M+2}g_{sc}(\lambda)}{d\lambda^{M+2}}\right)  \label{eq:rr.remainder}
\end{equation}
in which
\begin{equation}
\begin{split}
C_{M+2}&=(-1)^{M/2}\frac{1.3.5....(M+1)}{2^{(M+2)/2}(M+1)!}\\
&=\frac{(-1)^{M/2}}{2^{(M+1)}(M/2)!(M+2)} \label{eq:coef.m}
\end{split}
\end{equation}

 Here the definition of the coefficient $C_{M+2}$ differs (slightly) from that of reference \cite{2006m} by the quantity $(M+2)!$.
 \newline Since the Taylor series is truncated beyond the $(M + 2)^{th}$ term , all higher order terms are ignored. Consequently, taking them into account, the exact rest is obtained by summation of all the remaining elements. The actual remainder, that is,  the complete form of the remainder, can be written as an infinite sum:
 \begin{equation}
 R_{M+2,\gamma}(\lambda)=\sum_{k=M}^{\infty}\ C_{k+2}\frac{\gamma^{k+2}}{g_{sc}(\lambda)}\ \left(\frac{d^{k+2}g_{sc}(\lambda)}{d\lambda^{k+2}}\right)  \label{eq:r.remainder}
 \end{equation}
with $k=M, M+2, M+4,...\infty$, $M$ being even.
In this sum the coefficient $C_{k+2}$ is defined by:
 \begin{equation}
 C_{k+2}= \frac{(-1)^{M/2}}{(M/2)!}\frac{1}{\left[(k-M)/2\right]!\ 2^{k+1}(k+2)}
 \label{coff}
 \end{equation}
  The first element  (i.e., $k=M$) of this sum  (\ref{eq:r.remainder}) matches the one of  Eq. (\ref{eq:rr.remainder})  and the coefficient given by Eq. (\ref{coff})  reduces to the one of  Eq. (\ref{eq:coef.m}). \newline  It should be emphasized that Eq. (\ref{rel})  was obtained using the smoothing condition (\ref{eq:smootc}) and the classical limit (\ref{scl}), namely by: 
$\gamma\gtrsim\hbar\omega$  and $\lambda-V_{0}\gg\hbar\omega$.
\newline  Thus, (Eq. \ref{rel}) shows that the Strutinsky level density is only an approximation of the semi-classical one.
\newline Thus, the true value in such
calculations is the one given by the semi-classical method. Consequently, the
only question that arises is what accuracy can be obtained from the
Strutinsky method?\newline
In this respect, it seems better to deduce the "true" shell correction in a direct way, i.e. straightforwardly by
semi-classical methods. Indeed,  it is always possible, at the cost of extremely
complicated calculations, to use straight semi-classical formulas.
Nevertheless, the advantage of the Strutinsky method lies in
avoiding these complications. In this respect, in the Strutinsky method, we
only needs the knowledge of the level density deduced from the set of the
single particle energy levels and continuum. The semi-classical approximation is implicitly
contained in this quantity. As previously stated, the only problem is how to
deal with the Strutinsky method in order to obtain the best possible
accuracy, i.e. to make the Strutinsky density as close as possible to the semi-classical density. To this end, first of all, one has to minimize the remainder (see discussion in the next subsection below).

\subsection{The essential condition for precision: The asymptotic limit}

The remainder appears in Eq. (\ref{rel}). In order to make Strutinsky's density as close as possible to the semi-classical density, that is to obtain a good accuracy we must have: $Remainder\ll1$.


The order of magnitude of the remainder is given by Eq. (\ref{eq:rr.remainder}). Consequently, to obtain a good accuracy, we must impose:

\begin{equation}
\frac{\gamma^{M+2}}{g_{sc}(\lambda)/g^{(M+2)}_{sc}(\lambda)} \ll \frac{1}{C_{M+2}}
\label{fr}
\end{equation}

However, since $(1/C_{M+2})>1$,  Eq. (\ref{fr}) will be realized a fortiori, if:
\begin{equation}
\frac{\gamma^{M+2}}  {g_{sc}(\lambda)/g^{(M+2)}_{sc}(\lambda)} \ll1  
 \label{lessss}
\end{equation}

Furthermore,  we will also assume that the denominator of the above equation, is an increasing function of $ \lambda$.  Therefore, a necessary condition for the left member of Eq. (\ref{lessss}) to tend towards zero is that:
\begin{equation}
\frac{\gamma}{\lambda}\ll1 
\label{bottom}
\end{equation}

Before going further, to get a good idea, let us apply the formula given by equation (\ref{fr}) to the case of the cubic box with perfect reflecting walls seen in reference \cite{2006m}. Taking into account only the dominant term (Thomas-Fermi approximation) in the semi-classical level density, in this case, we will have:
\begin{equation}
 g_{sc}(\lambda)=Constant \times\sqrt{\lambda}
 \end{equation}
  Choosing for example  M=4, we obtain:


\begin{equation}
\left (\frac{\gamma}{\lambda}\right)^{4}\ll\frac{512}{15}
\end{equation}
which means that we will obtain a good accuracy for: $(\gamma/\lambda)\ll2.4$ and which is a less restrictive condition than the general case of the R.H.S of Eq. (\ref{lessss}). \newline
It is worth to note that, in formula (\ref{bottom}), the Fermi level for finite wells must be measured from the bottom of the well $V_{0}$, so:
\begin{equation}
\lambda-V_{0}\gg\gamma
\label{alim}
\end{equation}
Eq. (\ref{alim}) is called the ``asymptotic limit'' and is the necessary condition of the accuracy of the Strutinsky method. This formula is very similar to the one of the classical limit given by Eq. (\ref{scl}): $\lambda-V_{0}\gg\hbar\omega$. \newline
 Because the first energy level $\epsilon_{0}$ of the spectrum is close to the bottom of the wells $V_{0}$, the latter can be replaced by $\epsilon_{0}$ as in Ref. \cite{2006m}.

The basic assumption of the Strutinsky method is given by the smoothing condition of  Eq. (\ref{eq:smootc}). This condition means that the parameter $\gamma$  must be slightly larger than the inter shell spacing.  However, if we are only looking for smoothing, any value larger than the mean inter-shell spacing (including the very large $\gamma$ values) will achieve this smoothing.  Therefore, a priory, from the point of view of smoothing only, in principle, we must require a less restrictive condition :
\begin{equation}
\gamma\geq\hbar\omega
\label{eq:onlysmooth}
\end{equation}
The larger the $\gamma$, the stronger the smoothing.
This means that the usual condition (\ref{eq:smootc}), i.e. $\gamma\succsim\hbar\omega$,  contains ``by chance'' more information than smoothing alone. But, the reason why the smoothing parameter would be as small as possible is imposed by Eq. (\ref{alim}). In effect, in the latter, the only free parameter is $\gamma$,  so that, it is necessary to take the smallest $\gamma$ value compatible with the smoothing condition (\ref{eq:onlysmooth}). In practice, $\gamma$ value must be slightly larger than $\hbar\omega$. Usually, the standard choice is about:
\begin{equation}
\gamma\approx(1.0\sim2.0)\hbar\omega 
\label{cplat}
\end{equation}

 It is therefore the relationship between Strutinsky's method and the semi-classical method that legitimizes taking the condition of smoothing such as:
 \begin{equation}
 \gamma\succsim\hbar\omega
 \label{smoo}
 \end{equation}
 The latter is rewritten for the sake of clarity.

The asymptotic limit (\ref{alim}) and the smoothing condition (\ref{smoo}) can be summarized by a double equation:
\begin{equation}
\lambda-V_{0}\gg\gamma\gtrsim\hbar\omega
\label{doubcond}
\end{equation}
It should be noted that the semi-classical limit (\ref{scl}) is also contained in this double inequality. The latter plays an essential role in the fundamental formula (\ref{rel}). \newline 
Once again,  it is worthwhile to repeat that in the double condition, the only free parameter is $\gamma$ and it  must be chosen as small as possible from the `` point of view'' of the asymptotic limit. But its minimal value is limited by the smoothing condition. Therefore, it is not possible to ``play'' with its value as we want. In  fact the accuracy of the Strutinsky method is first  limited by the semi-classical limit: $\lambda-V_{0}\gg\hbar\omega$ which for light nuclei, is not fully fulfilled. This explains why this method is better for large nuclei, that is for which the Fermi level is large.

\subsection{The Strutinsky energy}
I point out that similar formulas to those of the level density can be
obtained for the energy. A connection between the energy of the Strutinsky
method and the semi-classical energy is then:

\begin{equation}
\overline{E}_{M,\text{ }\gamma}(\lambda)\approx
E_{sc}\left(\lambda\right)\left\{
1+\rho_{_{M+2},\gamma}\left(\lambda\right)\right\}    \label{eq:strutenerg} 
\end{equation}
with
\begin{equation}
E_{sc}(\lambda)=\int_{-\infty}^{\lambda}\epsilon g_{sc}(\epsilon)d\epsilon
\label{under}
\end{equation}

where the remainder of the energy is:
\begin{equation}
\rho_{_{M+2},\gamma}(\lambda)=
 \sum_{k=M}^{\infty}\frac{C_{k+2}\gamma^{k+2}}{E_{sc}(\lambda)}%
\int\limits _{-\infty}^{\lambda}\epsilon\left(\dfrac{d^{k+2}g_{sc}(\epsilon)%
}{d\epsilon^{k+2}}\right)d\epsilon   \label{eq:roremainder}
\end{equation}
where the coefficient $C_{k+2}$ is defined by Eq. (\ref{coff}), with the double condition (\ref{doubcond}).

\subsection{The singular case of the harmonic oscillator}   \label{singular}
We saw in section \ref{cl}, that the semi-classical level density of the harmonic oscillator was a polynomial of degree two. Considering the shape of the rest in Eq. (\ref{eq:roremainder}), it is obvious that for $M\geq2$, all the derivatives contained in that sum cancel simultaneously. Therefore, the remainder (\ref{eq:roremainder}) also cancels out and gives a perfect result for this example, provided that the smoothing condition is met. In this case, a perfect plateau is obtained and the Strutinsky energy becomes equal to the semi-classical energy, for any value of the $\gamma$  parameter. But this case is very particular and there is no other case for which all the derivatives of the semiclassical density in the remainder cancel simultaneously. This is the reason why it is impossible to find a plateau as clear as that of the harmonic oscillator. It is then necessary to try to minimize the rest by a procedure other  than the plateau condition. 
\section{The limit of the remainder when the order M tends toward large values\label{sec:Connection-with-the}}
As we have seen from  Eq. (\ref{doubcond}), it is not possible to improve the precision of the method by using a value of the parameter $\gamma$  as small as one would like because it is limited by the smoothing condition of  Eq.    (\ref{smoo}). Nevertheless, in the expression of the main term of the remainder Eq. (\ref{eq:coef.m}) or the complete form of the remainder Eq. (\ref{eq:r.remainder}), we have products of the form $C_{M+2}\gamma ^{M+2}$. As it can
easily be seen from Eq. (\ref{eq:coef.m}), the coefficient $C_{M+2}$ tends
toward zero as $M$ increases to infinity. Therefore, for a fixed value of
the parameter $\gamma ,$ this product also tends to zero as $M$ increases to
infinity. It is then tempting to simply take large values of $M$ to increase
this accuracy. But actually, things are no so simple that they appear.
Indeed, it has been already noticed in section \ref{sec:sed 5} that if $M$
is increased, the real width of the smoothing functions decreases so that it
is necessary to enlarge again the parameter $\gamma $ in order to fulfill
the fundamental relation of the smoothing condition in  Eq. (\ref{smoo}).%
\newline
Thus, as $M$ increases, $C_{M+2}$ decreases but the smoothing conditions
implies that $\gamma $ has to be increased at the same time. Therefore, as $%
M $ increases, the limit of the product $C_{M+2}\gamma ^{M+2}$remains
unclear.\newline
Finally, we cannot conclude by saying that taking large values of $M$ improves the accuracy
of the Strutinsky method. This will be shown in the following section.

\begin{figure*}
\includegraphics[width=1.5\columnwidth,keepaspectratio]{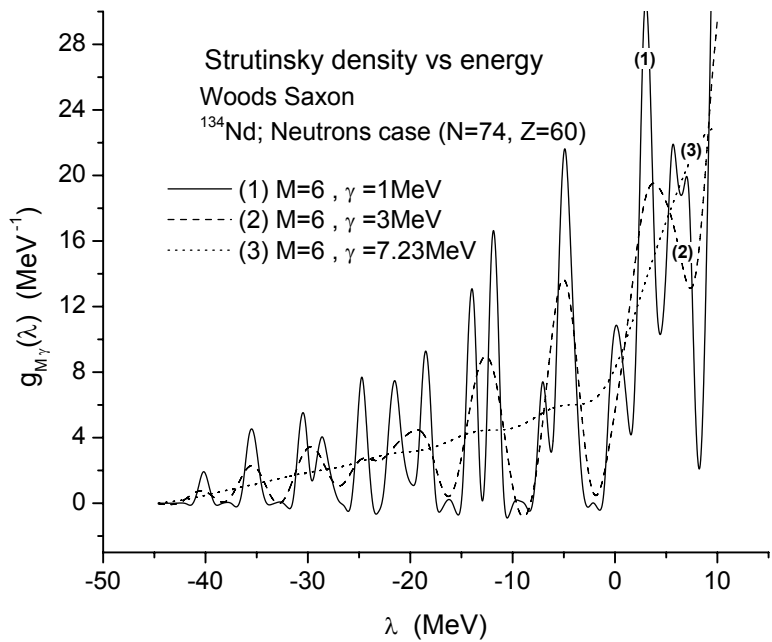}
\caption{Strutinsky level density as function of the Fermi level
for three values of the smoothing parameter. Note the oscillations (fluctuations) around the mean curve numbered by the number 3 between parentheses.}
\label{fig struto}
\end{figure*}


\section{The criterion of the monotonic behavior of the Strutinsky level density \label{sec:The-criterion-of}}
In Ref. \cite{1973b},  it has already been pointed out that the plateau condition cannot be achieved if the "average density contained in the quantum density" is not a polynomial and this point is perfectly explained in this paper (see subsection \ref{singular}). In this case, it has been suggested in reference \cite{1973b}  to replace the plateau condition by the stationariness condition which is an infinitesimal plateau (maximum or minimum), i.e. the points for which $\partial E/\partial\gamma=0$.
But the latter is marred by the fact that there are quite often several points of stationarity leading to ambiguity \cite{1973b}.  \newline Another problem in the old conception of the Strutinsky method, comes from the fact that, it has been found that even if the continuum is perfectly treated, the plateau is rarely encountered \cite{1998ve}. The latter suggests that an alternative recipe for defining shell correction for finite potentials, not based on the plateau condition but rather on the property of "quasi-linearity" of the smooth level density in the ``intermediate'' region (see, for example, Fig. \ref{fig ws}). In this respect, a new criterion based on the level density (instead the plateau condition) has already been proposed in that reference.  \newline
It turns out that the old conception of Strutinsky's method is unable to explain the origin of all these problems. In this paper, the relationship between Strutinsky's density and that of the semi-classical method explains that perfectly. Because of the remainder (see above), the smoothing parameter $\gamma$ must be "small enough" and because of the smoothing condition, it must "be sufficient" large. It is in the compromise between these two conditions that the optimization of Strutinsky's method must be carried out. Thus, the Strutinsky method is no more than a problem of optimization. \newline In any case, even in perfect optimization, the precision of Strutinsky's method remains conditioned by Eq. (\ref{scl}). The latter is not well realized for light nuclei. Once again, this paper explains why this method should be avoided for very light nuclei. \newline   It is within the natural framework of the relationship of the Strutinsky method with the semi-classical method that I propose a new criterion instead of the plateau condition. \newline
The most obvious thing we know is that, the rest given by Eq. (\ref{eq:r.remainder})  must be as small as possible. In this remainder, the only two free parameters are the smoothing parameter and the order of the curvature correction. From the previous section, we also know that it is not sure to improve the accuracy of this remainder by increasing the order $M$ because increasing $M$ implies to increase $\gamma$. It therefore remains "to play" with the only parameter $\gamma$. According to the double condition (\ref{doubcond}), this parameter must not be too large (because of its presence in the rest) or too small (because the smoothing might not be achieved).
In practice, the mean shell spacing $\hbar\omega$ is not accurately known a priori. One only knows that it is  of the order of $\hbar\omega\approx 41 A^{-1/3}$.  In fact, the  question is: What is the minimal  value of that parameter?  In other words, when can we pretend that the smoothing is actually realized? What is exactly the practical criterion for answering this question? Without beating about the bush, let's immediately answer this question (before justifying it later): In the following, we will see that smoothing is achieved when the Strutinsky density becomes monotonously increasing.

To demonstrate that, it is necessary to show how the Strutinsky level
density changes when the parameter $\gamma$ is increased from zero to
larger values, for a given order M (arbitrarily fixed). \newline
As function of the Fermi energy and for small values of $\gamma$, the
Strutinsky level density given by Eq. (\ref{eq:strutg}) is characterized by
an important oscillatory behavior around a  mean curve.
This is due to the smoothing functions (Eq. (\ref{eq:fm}))  which are close to
$\delta$ functions when their width is small. As the parameter $\gamma $
continues to increase, these oscillations decrease in amplitude and the
curve becomes more and more regular approaching thus this mean value for
which the oscillations disappear. In this respect, one can guess that the
disappearance of these oscillations marks the beginning of the smoothing and
fixes the optimal value of $\gamma$. Because by construction the smoothing
is realized for the smallest value of the parameter $\gamma$, i.e. for 
which the asymptotic limit (\ref{alim}) is best achieved, one deduces that the Strutinsky
level density will necessary be the closest to the semi-classical density. If we continue to increase this parameter we deviate from the condition (\ref{alim}) and therefore, we lose accuracy.
In practice, if we continue to increase the parameter $\gamma$, the curve remains regular, but begins to 
collapse more and more, starting from the top. 
One will
then move further and further away from this optimal value.\newline
A practical example is given in Fig. \ref{fig struto} . I have drawn the
Strutinsky level density for a fixed value of the order $(M=6)$, and three
values of the parameter $\gamma $. For the smallest value $\gamma =1Mev$,
one can see strong oscillations. These oscillations decrease in amplitude
for $\gamma =3MeV$ and disappear when the value of $\gamma $ reaches $
7.23MeV $ which is the optimal value, i.e. for which the curve becomes monotonous. These calculations are deduced from
the neutron case of $_{60}^{134}Nd_{74}$ . To obtain the set of energy
levels I have solved a realistic Schrodinger equation based on Woods-Saxon
potential following the method given in Ref.  \cite{2014mazizi}.\newline 
Thus, the optimal value of the $\gamma$ parameter is the smallest that makes the curve monotonous, without oscillations. In this respect, it is worth recalling that semi-classical density is a strictly increasing function.
 So, since the Strutinsky density is only an approximation of that one of the semi-classical method, it is natural to impose the same
property to the one calculated by the Strutinsky method. Thus, the
oscillatory behavior of the Strutinsky level density ceases when this
function becomes monotonically increasing. In practice, it is then enough to
gradually increase the parameter $\gamma $ and to verify from which value of
this parameter, the curve representing the Strutinsky level density becomes
monotonically increasing, without oscillations. One must not go beyond this
optimal value.

\section{According to the new criterion, do the results depend on the order M? \label{interrogation}}
In fact, in the previous section, the criterion of the monotonic behavior is
based only on the smoothing parameter $\gamma $. This criterion does not specify the
value of the order $M$. It only says that its value must be first
arbitrarily fixed. In fact, we know from section \ref{sec:sed 5} that the smoothing value of the parameter $\gamma$ depends on the order $M$. Thus, the optimization is done when $M$ is fixed. The question which arises to the mind is: if we choose
different values for the order $M$, can we obtain the same value for the
Strutinsky level density (and therefore for the Strutinsky energy and shell
correction)? \newline
In fact, we also  know that the remainder depends on $M$ and that $M$ varies by two units ($\Delta M=2$). It is to be expected that the relative variation of the remainder is smaller for the very large values of $M$, because in this case $M$ can be considered as a continuous variable. Consequently, for a question of  stability of results, it is advisable to take the largest possible values of $M$. Unfortunately, in this case rounding errors can become important and a compromise must be found.
This stability of the results is confirmed in the following numerical tests. \newline
In tables \ref{inf}, \ref{jnf} and \ref{knf}, I compare the Strutinsky calculations of energy with
the ones of the semi-classical method for three isotopes of $Neodymium$ . In
these calculations, the Strutinsky energy has been evaluated for different
values of the order $M$ using the present criterion of the monotonic behavior
to determine the smoothing value of the parameter $\gamma$. The
semi-classical energy is calculated by using the method given in Ref. \cite{2010ma2}. As already mentioned before, the Strutinsky calculations are
realized following my FORTRAN program given in Ref. \cite{2007ma}. I use
the Woods-Saxon potential with the universal parameters of Ref.  \cite{1987cw}.
\newline So, in these three cases, we can see
that the value $\gamma _{smooth}$ (of the parameter $\gamma $ ) which
initiate the smoothing (of the Strutinsky level density) increases with the
order $M$ but the Strutinsky energy, remains practically the same, close to the semi-classical energy. The bad value of the energy for $M=0$ is due to the absence of the curvature correction.
Thus, apart from the value $M=0$, the Strutinsky (smooth)
energy is found to be very close to the one deduced from the semi-classical
method. The relative error is about $0.0005MeV$ when one considers that the numerical value of the
semi-classical method is exact. In fact, the latter is also deduced numerically and also involves uncertainty.\newline 
From the tables, it is very clear that where the smoothing is realized, the Strutinsky
energy varies very little with $M$. Thus apart from $M=0$, the maximum
deviation (Min-Max difference for $M$ going from $M=2$ to $M=30$) are of
about $0.3MeV$,$0.6MeV$, $1.0MeV$ respectively for $N=60,70,80$ (the three
isotopes of $Nd$). From $M=16$ to $M=30$, the values of the Strutinsky
energy are practically constant to five significant digits in theses
examples. Thus, with these examples, one can expect stable results as soon
as we use medium values of $M$ ($\gtrsim 16$).\newline
For the lowest values of $M$ ($M=2-10$), the variations of the Strutinsky energy are somewhat larger. So, even among these values, there are some
which are very close to the semi-classical value. Obviously, the use of the Strutinsky method assumes that semi-classical energy is not known, otherwise it is of no interest. For this reason, it is however difficult to choose
a priori between them. Due to the "unavoidable remainder", there is always a
(small) uncertainty in this method. This uncertainty is specific in each case since the level density changes for each nucleus. Therefore a very perfect result would
only be due to chance.\newline
I can conclude this section by saying that as soon as the order $M$ is
fixed and the smoothing is realized for that value of $M$, the Strutinsky
energy becomes close to the semi-classical level density and insensitive to
the value of $M$, provided that $M$ is large enough (about $M\gtrsim 12)$.


\begin{table}
\begin{center}
\begin{tabular}{llll}
\hline \\[-1.8ex]
\multicolumn{4}{l}{$^{120}Nd(N=60,Z=60)$}\tabularnewline
\\[-1.8ex]\hline \\[-1.8ex]
$M$  & $\gamma_{smooth}$  & $E_{strut}$  & $E_{sc}$ \tabularnewline
 & $\left(MeV\right)$  & $\left(MeV\right)$  & $\left(MeV\right)$ \tabularnewline
 \\[-1.8ex]\hline \\[-1.8ex] 
0  & 4.88  & -1606.22  & -1583.25 \tabularnewline
2  & 6.21  & -1583.32  & -1583.25 \tabularnewline
4  & 7.1  & -1583.57  & -1583.25 \tabularnewline
6  & 7.86  & -1583.36  & -1583.25 \tabularnewline
8  & 8.64  & -1583.37  & -1583.25 \tabularnewline
10  & 9.31  & -1583.44  & -1583.25 \tabularnewline
12  & 9.81  & -1583.51  & -1583.25 \tabularnewline
14  & 10.42  & -1583.55  & -1583.25 \tabularnewline
16  & 10.91  & -1583.57  & -1583.25 \tabularnewline
18  & 11.38  & -1583.57  & -1583.25 \tabularnewline
20  & 11.86  & -1583.57  & -1583.25 \tabularnewline
22  & 12.32  & -1583.58  & -1583.25 \tabularnewline
24  & 12.76  & -1583.58  & -1583.25 \tabularnewline
26  & 13.18  & -1583.57  & -1583.25 \tabularnewline
28  & 13.59  & -1583.57  & -1583.25 \tabularnewline
30  & 13.98  & -1583.56  & -1583.25\tabularnewline
\end{tabular}
\caption{Comparison between the Strutinsky energy and the semi-classical energy for different order M of the curvature correction in the Neutrons' case of Neodymium 120 . The value of the smoothing parameter is given in Column 2.}
\label{inf}
\end{center}
\end{table}
\begin{table}
\begin{center}
\begin{tabular}{llll}
\hline \\[-1.8ex]
\multicolumn{4}{l}{$^{130}Nd(N=70,Z=60)$}\tabularnewline
\\[-1.8ex]\hline \\[-1.8ex]
$M$  & $\gamma_{smooth}$  & $E_{strut}$  & $E_{sc}$ \tabularnewline
 & $\left(MeV\right)$  & $\left(MeV\right)$  & $\left(MeV\right)$ \tabularnewline
  \\[-1.8ex]\hline \\[-1.8ex] 
0  & 4.58  & -1626.89 & -1603.20 \tabularnewline
2  & 5.85  & -1603.60 & -1603.20 \tabularnewline
4  & 6.70 & -1604.18 & -1603.20 \tabularnewline
6  & 7.39 & -1603.95 & -1603.20 \tabularnewline
8  & 8.11  & -1603.84 & -1603.20 \tabularnewline
10  & 8.74 & -1603.85 & -1603.20 \tabularnewline
12  & 9.30 & -1603.91 & -1603.20 \tabularnewline
14  & 9.81 & -1603.99 & -1603.20 \tabularnewline
16  & 10.29  & -1604.06 & -1603.20 \tabularnewline
18  & 10.74  & -1604.11 & -1603.20 \tabularnewline
20  & 11.17  & -1604.15 & -1603.20 \tabularnewline
22  & 12.32  & -1604.18 & -1603.20 \tabularnewline
24  & 11.59  & -1604.20 & -1603.20 \tabularnewline
26  & 12.00  & -1604.21 & -1603.20 \tabularnewline
28  & 12.78  & -1604.22 & -1603.20 \tabularnewline
30  & 13.15  & -1604.23 & -1603.20 \tabularnewline
\end{tabular}
\caption{\label{tab:table 2}Same as table I for Neodymium 130.}
\label{jnf}
\end{center}
\end{table}
\begin{table}
\begin{center}
\begin{tabular}{llll}
\hline \\[-1.8ex]
\multicolumn{4}{l}{$^{140}Nd(N=80,Z=60)$}\tabularnewline
\\[-1.8ex]\hline \\[-1.8ex]
$M$  & $\gamma_{smooth}$  & $E_{strut}$  & $E_{sc}$ \tabularnewline
 & $\left(MeV\right)$  & $\left(MeV\right)$  & $\left(MeV\right)$ \tabularnewline
  \\[-1.8ex]\hline \\[-1.8ex] 
0  & 4.32 & -1622.56 & -1598.14\tabularnewline
2  & 5.53 & -1598.46 & -1598.14\tabularnewline
4  & 6.33 & -1599.47 & -1598.14\tabularnewline
6  & 7.00 & -1599.23 & -1598.14\tabularnewline
8  & 7.65 & -1598.97 & -1598.14\tabularnewline
10  & 8.25 & -1598.82 & -1598.14\tabularnewline
12  & 8.78 & -1598.77 & -1598.14\tabularnewline
14  & 9.28 & -1598.75 & -1598.14\tabularnewline
16  & 9.75 & -1598.74 & -1598.14\tabularnewline
18  & 10.19 & -1598.75 & -1598.14\tabularnewline
20  & 10.61 & -1598.75 & -1598.14\tabularnewline
22  & 11.02 & -1598.74 & -1598.14\tabularnewline
24  & 11.42 & -1598.73 & -1598.14\tabularnewline
26  & 11.80 & -1598.72 & -1598.14\tabularnewline
28  & 12.18 & -1598.70 & -1598.14\tabularnewline
30  & 12.54 & -1598.68 & -1598.14\tabularnewline
\end{tabular}
\caption{\label{tab:table 3}Same as table I for Neodymium 140.}
\label{knf}
\end{center}
\end{table}

\section{Criticism of the plateau condition. Its true meaning}
 Let us examine the problem of the plateau condition \newline
The results of the Strutinsky  method strongly depend on the smoothing parameter $\gamma$  if the latter is chosen arbitrarily. In the old justification, it was argued that there is a range of $\gamma$ values for which the shell correction no longer depends on these values \cite{1967s,1968s}. It is the well known plateau condition. In fact, apart from the case of the  harmonic oscillator, there is no perfect plateau (see subsection \ref{singular}). Consequently, it is actually difficult to define exactly what a plateau is. A plateau is characterized by a perfect horizontal line. In practice, in the case of the harmonic oscillator, the length of the plateau depends on the size of the base as well as on the order $M$ of the curvature correction . In general, in real cases, we have curves with more or less pronounced slopes. The length of the plateau is also an ambiguous question since there are mini or micro plateaus that practically reduce to points. Sometimes the curve shows "steps" and we no longer know which is the "good" plateau. Thus, apart from the harmonic oscillator case, the plateau is an intuitive notion that is in itself tainted with ambiguities. 

The plateau condition is very rarely met \cite{1998ve}, and in a number of cases it is even non-existent . In order to remedy to these situations, it has been proposed to replace that criterion by the stationariness condition \cite{1973b} which in fact, is an infinitesimal plateau (that reduces to a single point).  An other criterion of optimization on the level density has been proposed in  Ref.  \cite{2000ve}.
\newline  In the light of the relation of the Strutinsky method with the semi-classical method, I will explain the plateau condition by arguments other than those which are usually given. My analysis will be illustrated by Fig. \ref{fig platx}.  \newline This figure shows the behavior of Strutinsky's energy as a function of the parameter $\gamma$. The horizontal line gives the value of the semi-classical density for the case in question. We can note the following points:

\begin{figure*}
\includegraphics[width=1.5\columnwidth,keepaspectratio]{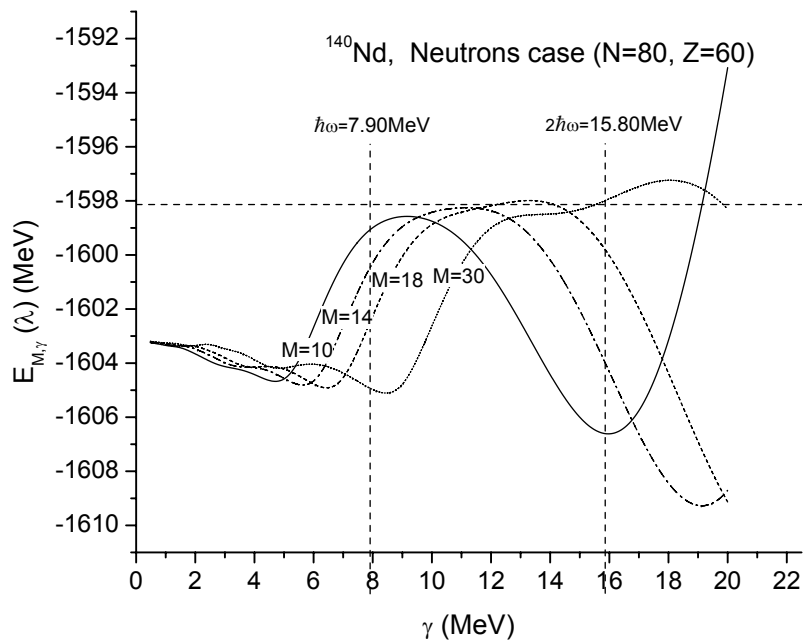}
\caption{Plateau condition for different orders of the curvature correction. Dashed horizontal line gives the value of the semi classical energy.}
\label{fig platx}
\end{figure*}

 i) It can be seen that the four curves, corresponding to four different values of $M$, start from the same value. This is easily explained since for $\gamma=0$, the Strutinsky density reduces to the quantum density and the Strutinsky energy becomes the sum of single particle states.  \newline   ii) We can also see that for each curve there are several extrema and the difficulty is to choose the right extremum. The usual recipe  is to choose the values in the range between $\hbar\omega$ and  $2\hbar\omega$ with the standard value given by Eq. (\ref{realistic}). In addition, the exrema of the four curves in the vicinity of the horizontal line (semi-classical value) do not have strictly the same value.\newline iii ) We also see that the extrema are shifted to the right as the value of $M$ increases. This has been well explained in section \ref{sec:sed 5}. This remark makes the recipe of choosing extrema in the interval $[\hbar\omega,2\hbar\omega]$ true only for the usual values of $M$ ($M$ less than 12). For the very large values of $M$, it is difficult to find this interval a priori. \newline  iv) For the curve with $M=30$, we see a plateau that is quite obvious. This is due to the fact that the semi-classical density is rather well approached by a polynomial of degree 30. Thus, It seems that sometimes the point  found from an optimization  turns into a mini plateau when we take large values of the parameter $M$. But this does not change our way of seeing. We obtain a "range" of values that gives the same energy.  \newline  Indeed, despite obtaining a plateau at $M=30$, , there is a difference between the energy value for this plateau and the exact semi-classical value. This is in line with what has been stated in this paper, namely that there is always a residual uncertainty due to the remainder. So, although there may be a plateau, this does not mean that its value is the true value. This only means that an optimal value could be found.  Basically, this remains an optimization problem. The thesis developed in this paper does not deny "the plateau" but explains its ambiguities and its limits. 
 \newline  v) Because the same example has been \textquotedblleft
treated\textquotedblright\ by my criterion in table \ref{knf} and by the
stationariness condition in Fig. \ref{fig platx}, it is to possible to compare
between the both methods through this illustration. Contrarily to the
difficulties encountered by the stationarity condition, one can see that the
criterion of the monotonic behavior gives the same value for the Strutinsky
energy, with a deviation of about  $0.1MeV$, as soon as soon as $M
$ exceeds the value $10$, without ambiguities. \newline  To conclude, it must be said that the plateau condition (or the stationary condition) is ambiguous since we generally obtain several different extrema. We can also say that the "chances" of obtaining a semblance of a plateau increase with the order $M$.  However, in this case, contrary to the usual values, they cannot be given by  the standard  equation (\ref{cplat}). It becomes difficult to make calculations over a large range of $M$ values. On the other hand, taking large values $M$, involves too much calculation with possible large roundoff errors. Moreover, for obvious practical reasons, it is difficult to go beyond $M = 30$, in systematic calculations. \newline
For all these reasons, the criterion proposed here seems clearer to us.

\section{Differences between the two types of density and their explanations}
This paper deals with the relationship of Strutinsky's method with the semi-classical method. It has therefore been established that Strutinsky's density is ultimately only an approximation of the semi-classical density. This may not seem obvious, since the two densities may appear very different on some points which are easily explained in this paper. 
\begin{figure*}
\includegraphics[width=1.5\columnwidth,keepaspectratio]{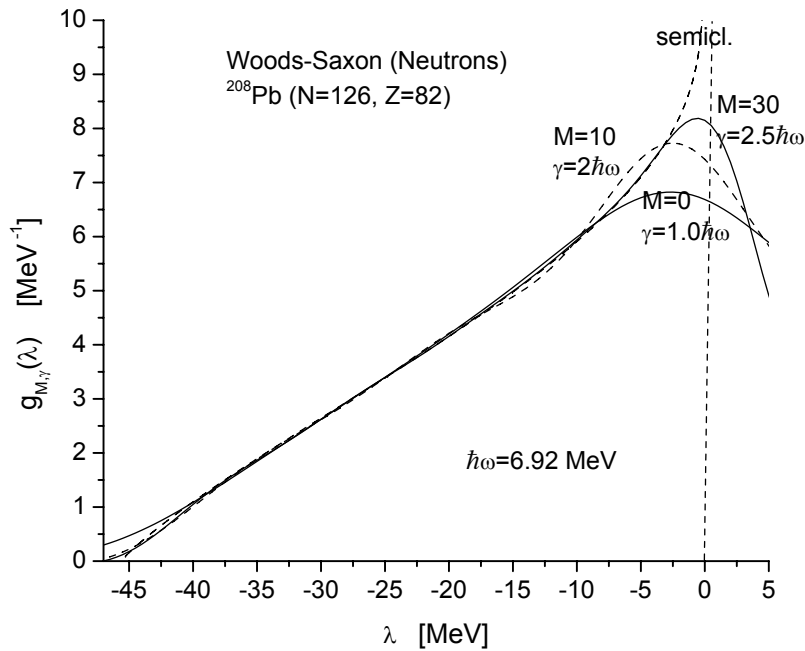}
\caption{Differences between the semi-classical and three Strutinsky densities at the top of of finite well for three distinct values of the order $M$. Note that in the vicinity of the singularity  $\lambda$=0 (vertical line), among the three curves, it is the one corresponding to $M=30$ that best approaches the semi-classical curve, more particularly the upper part of this curve. (From Ref. \cite{2006m}).}
\label{fig ws}
\end{figure*}
\begin{enumerate}[leftmargin=*]
\item It is well known that Strutinsky and semiclassical densities often give fairly similar results \cite{1975jbb}. However, this is not systematic (See also the special case of point 5 below). Indeed, as shown in equation \ref{rel}, the Strutinsky density and the semi-classical density are equal with an uncertainty due to the remainder. The latter, comes from the Taylor expansion of the semi-classical density (see  Eq. (\ref{666})) and contains all the Taylor's terms indexed by $m$ such as $m>M$  where  $M$ represents the order of the curvature correction. When the semi-classical density is a polynomial, Taylor's result is exact because the rest of the Taylor expansion  vanishes. This explains why in this case both methods give the rigorously the same result.
In the case where the semi-classical density is not a polynomial, there are two effects that are cumulated in the error of the Strutinsky method. The first is naturally the non-zero remainder of Taylor's expansion (up to the $M^ {th}$ term) of the semi classical level density. The second type of error ( inevitable) comes from the fact that the smoothing parameter $\gamma$ is the cause of another error that is added to the previous since it is contained in this rest (we must distinguish the rest of the Taylor expansion (Eq. (\ref{666})) from the rest of Strutinsky's method (Eq.  (\ref{rel}))  which is its integral). This means that this parameter must be as small as possible in order to make the remainder also as small as possible, so that the densities become as close as possible. However, this is not always achieved because, usually  this is not well understood. \newline
\item Another well established fact is that Strutinsky's method is all the more difficult to apply as the nuclei are lighter. But, this is easily explained. For light nuclei semi-classical limit condition  (\ref{scl}) , i.e. $\lambda-V_{0}\gg \hbar\omega$ is less true because the Fermi level is small or too small. Since Strutinsky density is only an approximation of semi classical density, both methods are less suitable for these nuclei.\newline
\item  In addition to the above point, it has been noticed in \cite{1998ve}, that the two densities differ for small $\lambda$ (see Fig. (\ref{fig ws})). In fact, for small $\lambda$ values,   the asymptotic limit (\ref{alim}), i.e. $\lambda-V_{0}\gg \gamma$,  is poorly verified. Remembering that the accuracy of the Strutinsky method is conditioned by the asymptotic limit, it is not surprising  to note slight differences for the lowest $\lambda$ values.  \newline
\item In the Strutinsky method, there are two free parameters: The smoothing parameter and the order of the curvature correction. The first is determined by the smoothing condition. However, the order $M$ corresponds to the degree of Taylor polynomial which approximates the semi-classical level density. Since a function is always better approached by a polynomial of a higher degree, there is a definite advantage in taking high $M$ values. The problem (see in particular section VII)  is that in Strutinsky's method, there is always a residual uncertainty (the remainder) that makes it impossible to obtain the desired (infinite) accuracy. For the Woods-Saxon potential, with the new criterion, stable numerical results (to five significant digits) seem to be assured for the range $ M = 16\sim30.$ (see tables I, II, III) \newline
\item Let's compare the two densities for finite wells. \newline
Since Strutinsky's density is only an approximation of the semi-classical density, no differences between these two densities should be observed. A minor difference is observed at the bottom of the the well and this  has already been explained just above. In fact, the main difference appears at the top of the well.  At this location, the semi-classical density has a singularity  ($\lambda=0$) for neutrons (or near the barrier for protons), while the Strutinsky density has only a peak  (see Fig. \ref{fig ws}). \newline  First of all, let's say that although the semi-classical density has a singularity, the value of the energy remains finite because it appears only under the integral symbol (see Eq. (\ref{under})). The density must not be considered as a function, but rather as a  distribution \cite{2010bh}. Moreover, It should also be noted that the demonstration which lead to the fundamental result  (\ref{rel})  does not allude to how to treat the continuum. It is only assumed that the discrete spectrum of eigenvalues and the continuum have been (correctly) resolved.  From then on, the two densities must be the same. Thus, Strutinsky's density should also tend towards infinity as we approach the threshold ($\lambda=0$). So how do we explain these differences? especially when the energy is close to the threshold ($\lambda=0$).  \newline
From Fig.\ref {fig ws},  it is clear that it is for the value $M=30$ (the largest) that the Strutinsky density best matches the semi-classical density at the top of the well, near the singularity $\lambda=0$. Since the order $M$ is no more than the degree of the Taylor polynomial  which approximates the semi-classical density, it is not surprising to obtain such results.  In effect, because of the singularity, the polynomial approximation is more difficult to achieve when one is close to this singularity (near $\lambda=0$). This means that the degree of the Taylor polynomial must be large enough as the Fermi level approaches the threshold energy. Thus, the difference between the Strutinsky density and the semi-classical density near the threshold energy can be corrected by taking large values for order $M$ (in order to improve the Taylor remainder and thereby the remainder of the Strutinsky method). Thus, for nuclei close to drip- lines, it is necessary to take fairly large values of $M$. It is possible to accept calculations with $M=10-14$ for standard situations  (Fermi level far from the singularity), but for nuclei near the threshold energy, i.e. near the drip lines, the previous values will not be sufficient. This error, has nothing to do with the continuum (assuming that it has been properly processed) as it is often claimed. \newline

\item Because Strutinsky's density is only an approximation of that of the semi-classical method,  it is initially more logical to work with semi-classical density. But in this method, the calculations are numerically much more cumbersome.
Furthermore, in most cases, the Strutinsky method applies after numerical resolution of the Schrodinger equation.  In this situation, it would be tedious to redo separately the resolution of the semi-classical part.
Although Strutinsky's method is only an approximation, the numerical aspect is much simpler with a lower risk of numerical error. In most cases the two methods give very similar results and the uncertainty due to the rest is generally very small. Finally, it can be said that Strutinsky's method remains competitive up to now.

\end{enumerate}

\section{Conclusion}
This paper is an extension of the previous one \cite{2006m}. It allowed me to review the Strutinsky method from a completely new angle. In this respect, a new criterion is proposed instead of the  one of the plateau condition. This criterion is more reliable and free from difficulties and  ambiguities. In addition, this study also explains in particular, why the Strutinsky method and the semi-classical approach lead in some situations to very similar results while in others, they give rise to disagreements (see the previous section). \newline
Despite the large number of remarks and details given in this paper, for the sake of clarity, the main results established in this study can be briefly summarized below in five points: The first three, recall the essential elements necessary to understand the Strutinsky method.  The next two explain how to smooth the level density and how to put it into practice by applying the new criterion. These points are as follows:
\begin{enumerate}[leftmargin=*]
\item The Strutinsky method is only an approximation of the semi classical method and
results from the compromise between two antagonistic conditions which are
the smoothing condition and the asymptotic limit.
\item By construction, the Strutinsky level density is a function that presents
oscillations. These oscillations occur around an average curve. It turns out that, the average curve is an approximation of the semi classical level density. The difference between them constitutes the remainder. The latter remains small but cannot be canceled.
\item The Strutinsky method consists of adjusting the smoothing parameters 
$\gamma $ to smooth that level density, that is to say, to obtain the average curve cited above. Smoothing means making the oscillations disappear in order to obtain that average curve.
\item The "competition" between the condition of the asymptotic limit and the smoothing condition shows that the best choice is to take the smallest value of the parameter $\gamma 
$ that smoothes the Strutinsky level density. For this purpose, the
criterion of the monotonic behavior of the Strutinsky level density has been
adopted. It gives the optimal value of the parameter $\gamma$, without any ambiguity.  In practice, it is necessary to increase $\gamma $ little by little and to see from which value, the curve becomes strictly increasing, without oscillations. In this way, the average curve is reached. One must not go beyond this value because higher $\gamma$ values cause loss of accuracy ( the consequence is that the curve collapses).
\item For each value of the order $M$, one can find an optimal value of the
parameter $\gamma $. Once this has been made, the Strutinsky energy depends very
little on the order $M$, provided that the value of $M$  is not too small.
\end{enumerate}



\begin{thebibliography}{99}
\bibitem{1966s} V. M. Strutinsky, Yad. Fiz. 3, 614 (1966) {[}Sov. J. Nucl.
Phys. 3, 449 (1966){]}.

\bibitem{1967s} V. M. Strutinsky, Nucl. Phys. A95, 420 (1967).

\bibitem{1968s} V. M. Strutinsky, Nucl. Phys. A122, 1 (1968).
\bibitem{1972br} M. Brack, J. Damgaard, A.S. Jensen, H.C. Pauli, V.M.
Strutinsky and C. Y. Wong. Rev. Mod. Phys. 44, 320 (1972)

\bibitem{2018ivan} F. A. Ivanyuk, C. Ishizuka, M. D. Usang, And S. Chiba,
Phys. Rev. C 97, 054331 (2018)

\bibitem{2018qing} Qing-Zhen Chai, Wei-Juan Zhao, Min-Liang Liu And Hua-Lei
Wang , Chinese Physics C, Volume 42, Number 5 (2018)

\bibitem{2018dobrov} A. Dobrowolski, K. Mazurek, And A. G\~{A}
, Phys. Rev. C 97, 024321 ( 2018)
\bibitem{2018qingo} Qing-Zhen Chai Wei-Juan Zhao Hua-Lei Wangmin-Liang Liu
Fu-Rong Xu, Progress Of Theoretical And Experimental Physics, Volume 2018,
Issue 5 (2018) 053d02

\bibitem{1981braquen} M.Brack and P.Quentin, Nuclear Physics A, Volume 361,
Issue 1, (1981) pp. 35-82

\bibitem{2006m} Mohammed-Azizi B. and Medjadi D. E. Phys. Rev. C, 2006, v.
74, 054302.



\bibitem{1933k} J. G. Kirkwood, Phys. Rev. 44, 31 (1933).

\bibitem{1934w} E. P. Wigner, Phys. Rev. 46, 1002 (1934).
\bibitem{1973j} B. K. Jennings, Nucl. Phys. A207, 538 (1973)
\bibitem{1975j} B. K. Jennings and R.K. Bhaduri, Nucl. Phys. A237, 149
(1975).

\bibitem{2010bh} A. Bhagwat, X. Vinas, M. Centelles, P. Schuck and R. Wyss  Phys. Rev. C, 2010, v.
81, 044321. 

\bibitem{1975jbb} B. K. Jennings, R. K. Bhaduri, and M. Brack, Nucl. Phys.
A253, 29 (1975)






\bibitem{1973b} M. Brack and H. C. Pauli, Nucl. Phys. A207, 401 (1973).

\bibitem{1998ve} T. Vertse, A. T. Kruppa, R. J. Liotta, W. Nazarewicz,
N. Sandulescu, and T. R. Werner, Phys. Rev. C 57, 3089 (1998).





\bibitem{1976je} B. K. Jennings, Ph. D. Thesis, McMaster University, Hamilton, Ontario, 1976

\bibitem{2001ul} D. Ullmo, T. Nagano, S. Tomsovic, and H. U. Baranger, Phys. Rev. B, v. 63, 1972, 125339.




\bibitem{2000ve} T. Vertse, A. T. Kruppa, and W. Nazarewicz Phys. Rev. C, 2000, v.
61, 064317


\bibitem {2007ce}  M. Centelles, P. Schuck, and X. Vinas, Ann. Phys. 322, 363
(2007)










\bibitem{1949mayer} M. G. Mayer, Phys. Rev. 75, 1969 (1949)  



\bibitem{1949haxel} Otto Haxel, J. Hans D. Jensen, and Hans E. Suess Phys.
Rev. 75, 1766 (1949)




\bibitem{1972b} M. Bolsterli, E. O. Fiset, J. R. Nix, and J. L. Norton,
Phys. Rev. C 5, 1050 (1972).


\bibitem{2014mazizi} Mohammed-Azizi B. and Medjadi D. E. Comput. Phys.
Commun., 2014, v. 185, 3067-3068


\bibitem{2010ma2} Mohammed-Azizi B. Intern. Journal of Modern Physics C,
2010, v. 21 (5), 681-694
\bibitem{2007ma} Mohammed-Azizi B. and Medjadi D. E. Comput. Phys. Commun.,
2007, v. 176, 634-635



\bibitem{1987cw} S. Cwiok, J. Dudek, W. Nazarewicz, T. Werner, Comput. Phys.
Commun. 46, 379 (1987).
















\end{thebibliography}
\end{document}